\documentstyle[epsf,aps]{revtex}

\begin{document}
\draft

\title{Statistical mechanics of random two-player games}
\author{J.~Berg \thanks{email: johannes.berg@physik.uni-magdeburg.de}}
\address{Institute for Theoretical Physics\\
         Otto-von-Guericke University, Postfach 4120, \\
         D-39016 Magdeburg, Germany }

\maketitle
 
\begin{abstract}
Using methods from the statistical mechanics of disordered systems we analyze the 
properties of bimatrix games with random payoffs in the limit where the number 
of pure strategies of each player tends to infinity. We analytically calculate quantities such 
as the number of equilibrium points, the expected payoff, and the fraction of strategies 
played with non-zero probability as a function of the correlation between the payoff 
matrices of both players and compare the results with numerical simulations. 
\end{abstract}

The adaptation to the behaviour of others and to a complex environment is a 
process central to economics, sociology, international relations, and politics. 
Game theory aims to model problems of strategic decision-making 
in mathematical terms: Two or more interacting participants, called players,    
make decisions in a competitive situation. Each player receives a reward, called the 
payoff, which not only depends on his own decision, but also on those of the other players. 
In the generic set-up a number of players choose between different
strategies, the combination of which determines the outcome of the
game specified by the payoff to each player. Each player strives to achieve as 
large a payoff as possible. One of the
cornerstones of modern economics and game theory is the
concept of a Nash equilibrium \cite{Nash}, for an introduction see also \cite{WJ}. 
A Nash equilibrium (NE)  
describes a situation where no player can unilaterally improve his
payoff by changing his individual strategy given that the other players 
all stick to their strategies. However this concept is thought to
suffer from the serious drawback that in most games there is a large 
number of Nash equilibria with different
characteristics but no means of telling which one will be chosen by
the players, as would be required of a predictive theory.

This conceptual problem already shows up in the paradigmatic model of
a bimatrix game between two players $X$ and $Y$ where player $X$
chooses a so-called \emph{pure strategy} $X_i \in (1 \ldots N)$ with 
probability $x_i \geq 0$ and player $Y$ chooses strategy $Y_j \in (1 \ldots N)$ with probability
$y_j \geq 0$. The vectors ${\bf x}=(x_1,...,x_N),\ {\bf
y}=(y_1,...,y_N)$ are called \emph{mixed strategies} and are constrained to
the $(N-1)$-dimensional simplex by normalization. For a pair of pure
strategies $(i,j)$ the payoff to player $X$ is given by the
corresponding entry in his payoff matrix $a_{ij}$ whereas the payoff
to player $Y$ is given by $b_{ij}$. The \emph{expected payoff} to player $X$
is thus given by $\nu_x({\bf x},{\bf y})=\sum_{i,j} x_i a_{ij} y_j$
and analogously for player $Y$. A Nash equilibrium $({\bf x}^*,{\bf y}^*)$ is
defined by
\begin{eqnarray}
\label{defNE}
  \nu^x({\bf x}^*,{\bf y}^*) & =& \mbox{max}_{\bf x}\ \nu^x({\bf
  x},{\bf y}^*) \nonumber\\ \nu^y({\bf x}^*,{\bf y}^*) & =&
  \mbox{max}_{\bf y}\ \nu^y({\bf x}^*,{\bf y}) \ .
\end{eqnarray}

The condition for a NE with a \emph{given set} of 
expected payoffs $\nu^x$ and $\nu^y$ 
may be written as 
\begin{eqnarray}
\label{eq:NE2}
 \sum_j a_{ij} y_j-\nu^x  \leq  0 & \ \ x_i\geq 0 &\ \  x_i (\sum_j a_{ij} y_j-\nu^x)=0 \ \ \forall i
 \nonumber\\
 \sum_i x_i b_{ij} -\nu^y  \leq  0 & \ \  y_j\geq 0 &\ \  y_j(\sum_i x_i b_{ij} -\nu^y)=0 \ \ \forall j \ , 
\end{eqnarray}
where we have dropped the $^*$ indices for simplicity. The first column ensures that there are no 
pure strategies (and thus also no mixed strategies) 
which will yield a payoff larger than $\nu^x$ to player $X$ and $\nu^y$ to 
player $Y$. Thus no player will have a reason to deviate from his mixed strategy. 
The second column ensures that the mixed strategies may be interpreted as 
probabilities and the third ensures that $\nu^x=\sum_{i,j} x_i a_{ij} y_j$ 
and analogously for player $Y$.  
In this situation there exists no mixed strategy which
will increase the expected payoff to $X$ if $Y$ does not alter his
strategy and vice versa for $Y$. Nash's theorem \cite{Nash} states
that for any bimatrix game there is at least one NE.

The third column in (\ref{eq:NE2}) states that whenever $x_i$ 
is strictly positive, $\sum_j a_{ij} y_j=\nu^x$ and whenever $\sum_j a_{ij} y_j-\nu^x$ is strictly 
negative, $x_i$ is zero. Thus for a given set of strategies played with non-zero probability 
(out of $4^N$ possible choices), 
the values of all non-zero components of a mixed strategy can be determined by 
solving the resulting linear equations  $\sum_j a_{ij} y_j=\nu^x \, \forall i:x_i>0$ and 
$\sum_i x_i b_{ij} =\nu^y \, \forall j:y_j>0$ subject to the normalization condition. 

Apart from applications in economics, politics, sociology, and
mathematical biology, there exists a wide body of mathematical
literature on bimatrix games concerned with fundamental topics such as
exact bounds for e.g. the number of NE \cite{Keiding} and efficient
algorithms for locating them \cite{Stengel}. For games even of
moderate size a large number of NE are found, forming a set of
disconnected points. In general the different NE all yield 
different expected payoffs to the players.

However many situations of interest are characterized by a large number
of possible strategies and complicated relations between the strategic
choices of the players and the resulting payoffs.  In such cases it is
tempting to model the payoffs by random matrices in order to calculate
{\it typical} properties of the game. This idea is frequently used in
the statistical mechanics approach to complex systems such as spin
glasses \cite{MPV,Young}, neural networks \cite{HKP}, evolutionary
models \cite{DieOpp,OppDie}, or hard optimization problems \cite{MezardParisi,MoZe}.
Recently this approach has been used to investigate the typical
properties of zero-sum games \cite{BergEngel} obeying 
$a_{ij}=-b_{ij}$. A partial analysis of bimatrix games using the so-called 
annealed approximation has been given in \cite{BergWeigt}. 

In this paper we investigate the properties of Nash
equilibria in bimatrix games with a large number of pure strategies and 
random entries of the payoff matrices. In this approach 
characteristics of the game are encoded in the
distribution of payoff matrices -- with only a few parameters --
instead of the payoff matrices themselves. Using
techniques from the statistical mechanics of disordered systems such as the 
replica-trick we calculate the typical number of NE with a given payoff. 

The paper is organized as follows: Having set up the probability distribution of 
payoffs to be considered, we construct an indicator function for NE which will allow us to 
count the number of NE. Then the average of the logarithm of the number of NE over the 
disorder will be calculated. The solution is discussed both in game theoretic and in 
geometric terms and is compared with the results of numerical simulations. Finally, 
we give a summary and an outlook to future developments. 

\section{The distribution of payoff matrices}
\label{ch3dist}

We consider bimatrix games with square payoff matrices $\{a_{ij},b_{ij} \}$ $i,j=1 \dots N$, where 
the thermodynamic limit consists of $N \to \infty$. We assume that the entries of 
the payoff matrices at different sites are identically and independently distributed.  
Since the two payoff matrices may be multiplied by any 
constant or have any constant added to them without changing the 
properties of the game in any material 
way there is no loss of generality involved in considering 
payoffs of order $N^{-1/2}$ and of zero mean. 
In the thermodynamic limit one finds that only the first 
two moments of the payoff distribution 
are relevant, as is generally the case in fully connected disordered systems 
described by mean-field theories. Hence the entries of the payoff matrices may be 
considered to be Gaussian distributed. Then the only property of 
the distribution of payoffs which 
is not fixed by these specifications is the correlation $\kappa$ 
between entries at the same site of the two payoff matrices. 

We thus choose the entries of the payoff matrices
to be drawn randomly according to the probability 
distribution
\begin{equation}
\label{eq:payensemble}
  p(\{a_{ij}\},\{b_{ij}\}) = \prod_{ij}
  \frac{N}{2\pi\sqrt{1-\kappa^2}}
  \exp\left\{-\frac{N (a_{ij}^2-2 \kappa a_{ij} b_{ij} + 
  b_{ij}^2 )}{2(1-\kappa^2)} \right\},
\end{equation}
i.e. a Gaussian distribution with zero mean, variance $1/N$ and correlation
${\langle \langle} a_{ij}b_{kl} {\rangle \rangle}=\kappa\delta_{ik}\delta_{jl}/N$ 
for all pairs $(i,j)$ and $(k,l)$.
Here and in the following, the double angles denote the average over the
payoff distribution (\ref{eq:payensemble}).
For $\kappa=-1$ (\ref{eq:payensemble}) includes a Dirac-delta $\delta(a_{ij}+b_{ij})$ corresponding 
to a zero-sum game and we recover the situation of \cite{BergEngel} as a special case. 
$\kappa=0$ corresponds to uncorrelated 
payoff matrices and $\kappa=1$ is the so-called symmetric case 
$a_{ij}=b_{ij}$ where the two players always receive identical payoffs. 
 
Thus the parameter $\kappa$ describes the degree of similarity between the payoffs to either player and 
can be used to continuously tune the game from a zero-sum game to a purely symmetric game.  
In the former case, the gain of one player is the loss of the other, so generally negative $\kappa$ 
correspond to a competitive situation, whereas for positive $\kappa$ there are many pairs of strategies 
which are beneficial to both players. 

\section{The entropy of Nash equilibria and the indicator function}
\label{ch3indi}

In this section we construct an indicator function which is zero at a NE with 
payoffs $\nu^x$ and $\nu^y$ to players $X$ and $Y$ respectively, and non-zero everywhere 
else. This function will be made the argument of a properly normalized Dirac-delta function. 
Integrating the Dirac-delta function over the mixed strategies of both players, we are effectively 
counting the number of NE with the specified payoffs. From the resulting spectrum of NE the 
statistical properties of NE in bimatrix games may be deduced. Since we expect the number of 
NE to scale exponentially with the size of the game our central tool of investigation will be the 
\emph{entropy of Nash equilibria} 
defined by $S(\nu^x,\nu^y)=\frac{1}{N}\ln {\cal N}(\nu^x,\nu^y)$ where 
${\cal N}(\nu^x,\nu^y)$ is the number of NE with the specified payoffs per unit interval within a  
small interval around $\nu^x$ and $\nu^y$. Since $NS(\nu^x,\nu^y)$ is 
expected to be an extensive quantity, we may assume that $S(\nu^x,\nu^y)$ is self-averaging and in 
the thermodynamic limit the average value of the entropy  
will be realized with probability one.  
Hence the central goal of our calculation will be to evaluate ${\langle \langle} S(\nu^x,\nu^y) {\rangle \rangle}$. 
 
In this framework the total number of NE is given by 
\begin{equation}
{\cal N}= \int d \nu^x d \nu^y e^{NS(\nu^x,\nu^y)} \ , 
\end{equation}
so in the thermodynamic limit the NE will be exponentially dominated by the maximum of the 
curve $S:=\max S(\nu^x,\nu^y)$. This implies that a randomly chosen NE will yield the 
payoffs where the maximum occurs with probability 1. 
On the other hand, the line where $S(\nu^x,\nu^y)=0$ 
delimits the smallest and the largest values of $\nu^x,\nu^y$ for which there is still an exponential 
number of NE. 

The three expressions may be encoded in a single condition \cite{Opper} by 
introducing the variables ${\bf \tilde{x}}$ and ${\bf \tilde{y}}$ with 
\begin{eqnarray}
\tilde{x}_i&=&\left\{\begin{array}{ll} x_i & {x}_i > 0 \\
                                     \sum_j a_{ij} y_j -\nu^x &  {x}_i = 0 
                   \end{array} \right. \nonumber \\
\tilde{y}_j&=&\left\{\begin{array}{ll} y_j & y_j > 0 \\
                                     \sum_i x_i b_{ij} -\nu^y &  y_j = 0 
                   \end{array} \right. \ .
\end{eqnarray}
Condition (\ref{eq:NE2}) may be written as 
\begin{eqnarray}
I^x_i &=\tilde{x}_i \Theta(-\tilde{x}_i) - (\sum_j a_{ij}\tilde{y}_j \Theta(\tilde{y}_j)  -\nu^x) =& 0 \nonumber \\
I^y_j &=\tilde{y}_j \Theta(-\tilde{y}_j) - (\sum_j \tilde{x}_i \Theta(\tilde{x}_i) b_{ij} -\nu^y) =& 0 \ ,
\end{eqnarray}
so for positive $\tilde{x}_i$, $\tilde{x}_i=x_i$, whereas for negative $\tilde{x}_i$ we have 
$\tilde{x}_i=\sum_j a_{ij} \tilde{y}_j \Theta(\tilde{y}_j)  -\nu^x$. Furthermore we have $x_i=0$ if 
$\tilde{x}_i<0$ and $\sum_j a_{ij} \tilde{y}_j \Theta(\tilde{y}_j)  -\nu^x=0$ for $\tilde{x}_i>0$. 
The condition $x_i (\sum_j a_{ij} y_j-\nu^x)=0$ is thus satisfied automatically. Analogous relations 
hold for player $Y$. The new variables therefore serve as a 
convenient tool to encode the `complementary' quantities $x_i$ and $\sum_j a_{ij} y_j -\nu^x$ in a single 
variable. Analogous relations hold for player $Y$. 
The density of NE with payoffs $\nu^x$ and $\nu^y$ may thus be written as 
\begin{equation}
\label{ch3part}
{\cal N}(\nu^x,\nu^y)=\int d \mu({\bf \tilde{x}}) d\mu({\bf \tilde{y}}) 
   \prod_i \delta(I_i^x) \prod_j \delta(I_j^y) 
   \| \frac {\partial ({\bf I}^x,{\bf I}^y)}{\partial ({\bf \tilde{x}},{\bf \tilde{y}})} \| \ , 
\end{equation}
where the mixed strategies are rescaled to $\sum_i x_i=\sum_j y_j=N$ so we define the measure $d \mu$ as 
\begin{equation}
\label{ch3measure}
d\mu({\bf \tilde{x}}) = \prod_{i} d\tilde{x}_i 
\delta \left(\sum_i \tilde{x}_i \Theta(\tilde{x}_i) -N \right) \ . 
\end{equation}
This scaling of the mixed strategies assumes that the an extensive number of strategies are 
played with non-zero probability, so the individual terms $x_i$ and $y_j$ are all of order 
${\cal O}(1)$. 
The integrals over ${\bf \tilde{x}},{\bf \tilde{y}}$ effectively amount to choosing a set 
of strategies with $\tilde{x}_i>0$ 
and $\tilde{y}_j>0$, solving the resulting linear equations for the components played with 
non-zero probability, and checking if this candidate for a NE fulfills the 
conditions (\ref{eq:NE2}). It may thus be viewed as performing the so-called support 
enumeration algorithm analytically \cite{Stengel}. 

\section{Calculation of the typical number of Nash equilibria}
\label{ch3calc}

In this section we calculate the average of $S(\nu_x,\nu_y)$ over the disorder (\ref{eq:payensemble}). 
In order to represent the logarithm of (\ref{ch3part}) we use the replica-trick 
$\ln {\cal N} = \lim_{n \to 0} \frac{d}{dn} {\cal N}^n$ and compute the average over ${\cal N}^n$ 
for integer $n$ taking the limit $n \to 0$ by 
analytic continuation at the end. Using integral representations of the Dirac-delta function we 
obtain 
\begin{eqnarray}
\label{partrep1}
{\cal N}^n (\nu^x,\nu^y)  &=&
\prod_{a,i}^{n,N} \int \frac{d \mu({\bf \tilde{x}}^a) d \hat{x}_i^a}{2 \pi} \prod_{a,j} 
\int \frac{d \mu ({\bf \tilde{y}}^a) d \hat{y}^a_j}{2 \pi} \nonumber \\ 
&&\exp \{ -i\sum_{a,i} \tilde{x}^a_i \Theta(-\tilde{x}^a_i)\hat{x}^a_i 
+i\sum_{a,i,j}\hat{x}^a_i a_{ij}\tilde{y}^a_j \Theta(\tilde{y}^a_j)   
-i \nu^x \sum_{a,i} \hat{x}^a_i\nonumber \\ 
&&-i\sum_{a,j} \tilde{y}^a_j \Theta(-\tilde{y}^a_j)\hat{y}^a_j 
+ i\sum_{a,i,j} \tilde{x}^a_i \Theta(\tilde{x}^a_i)  b_{ij} \hat{y}^a_j 
-i \nu^y \sum_{a,j} \hat{y}^a_j   
   \} \nonumber \\
&&(\| \det(D) \|)^n \ ,
\end{eqnarray} 
where $a$ runs from $1$ to $n$. 
The most awkward term of this expression is the absolute value of the normalizing determinant, i.e. 
the Jacobian matrix of ${\bf I}^x,{\bf I}^y$ given by 
\begin{equation}
  D:= \frac {\partial ({\bf I}^x,{\bf I}^y)}{\partial ({\bf \tilde{x}},{\bf \tilde{y}})} =
  \left( 
  \begin{array}{ll} 
  \delta_{ii'}\Theta(-\tilde{x_i}) & -a_{ij}\Theta(\tilde{y_j}) \\
   -b_{ij}\Theta(\tilde{x_i})  &\delta_{jj'}\Theta(-\tilde{y_j})  \\
  \end{array} \right) \ .
\end{equation}   
which arises from the coefficients of $\tilde{x}$ and $\tilde{y}$ in ${\bf I}^x$ and ${\bf I}^y$. 
Since we are only interested in the absolute value of the determinant, we are free to interchange 
rows and columns of this matrix. Rearranging the rows and columns of $D$ such that the $p_x N$ 
strategies with $\tilde{x_i}>0$ and the $p_y N$ strategies with $\tilde{y_j}>0$ are grouped together, 
one finds that only the 
resulting quadratic submatrix of size $N(p_x+p_y)$ by $N(p_x+p_y)$ contributes to the determinant of $D$. 
From (\ref{eq:NE2}) one finds that the distinction between $p_x$ and $p_y$ is immaterial since the 
number of strategies played by player $X$ at any NE always equals that played by $Y$ and the 
determinant of $D$ is zero for $p_x \neq p_y$. In the following we will assume that $\frac{1}{N} \ln (\det D)$ 
is a self-averaging quantity depending on $p_x=p_y$.  
Splitting off the normalizing determinant, the average over the payoffs may now easily be performed, 
details of the calculations are given in appendix \ref{appendix}. The average over the disorder 
introduces a coupling between the replicas and one introduces the symmetric matrix of the overlaps 
between mixed strategies of each player as order parameters, 
\begin{equation}
\label{ch3opsq}
q_{ab}^x=\frac{1}{N} \sum_{i} \tilde{x}_i^a \Theta(\tilde{x}_i^a) \tilde{x}_i^b \Theta(\tilde{x}_i^b)  \ , \ \ \ \ \  
q_{ab}^y=\frac{1}{N} \sum_{j} \tilde{y}_j^a \Theta(\tilde{y}_j^a) \tilde{y}_j^b \Theta(\tilde{y}_j^b) \ , 
\end{equation}
as well as their conjugates $\hat{q}^{x,y}_{ab}$. 
At non-zero values of $\kappa$ we also obtain terms which couple the phase-space variables $\tilde{x}_i$ 
to the auxiliary variables $\hat{x}_i$ and similarly for player $Y$, so we also introduce the order 
parameters 
\begin{equation}
\label{ch3opsr}
R_{ab}^x=\frac{1}{N} \sum_{i} i \hat{x}^a_i \tilde{x}_i^b \Theta(\tilde{x}_i^b)  \ , \ \ \ \ \ 
R_{ab}^y=\frac{1}{N} \sum_{j} i \hat{y}^b_j \tilde{y}_j^a \Theta(\tilde{y}_j^a) \ .
\end{equation}
Similarly, in order to include the normalizing determinant we introduce the order parameters 
\begin{equation}
\label{ch3opsp}
p_{a}^x=\frac{1}{N} \sum_{i} \Theta(\tilde{x}_i^a) \ , \ \ \ \ \  
p_{a}^y=\frac{1}{N} \sum_{j} \Theta(\tilde{y}_j^a)
\end{equation}
giving the fraction of strategies played at a NE. 

Anticipating the limit $n \to 0$, the quenched average of the normalizing determinant may be 
computed using results from the theory of random matrices as outlined in appendix \ref{appendix} giving 
${\langle \langle} \ln(\| \det(D) \|) {\rangle \rangle} =Np( \ln p -1)$.

We finally obtain 
\begin{eqnarray}
\label{partrep2}
 {\langle \langle} {\cal N}^n (\nu^x,\nu^y){\rangle \rangle} &&  =
  \prod_{a \geq b}\int \frac{dq_{ab}^{x,y} d\hat{q}_{ab}^{x,y}}{2i \pi/N} 
  \prod_{a,b}\int \frac{dR_{ab}^x dR_{ab}^y} {2i \pi/(\kappa N)}
  \prod_a \int \frac{dp_{a}^{x,y} d\hat{p}_{a}^{x,y}}{2i \pi/N} \, \delta(p^x_a-p^y_a)    
  \prod_a \int \frac{dE_a^{x,y}}{2i\pi/N}  \nonumber \\
&& \exp \{ -N \sum_{a \geq b} q_{ab}^{x,y} \hat{q}_{ab}^{x,y} - \kappa N \sum_{a,b} R_{ab}^x R_{ab}^y + 
  N \sum_a p^{x,y}_a \hat{p}^{x,y}_a  + N\sum_a E^{x,y}_a \} \nonumber \\
&& \exp \{ N \left[ G^x +  G^y \right] \} {\langle \langle} \| \det(D) \|^n {\rangle \rangle} \ ,
\end{eqnarray}
where 
\begin{eqnarray}
\label{Gdef} G^x&=&\ln \prod_a \int \frac{d \tilde{x}^a d \hat{x}^a}{2 \pi} 
\exp\{{\cal L}^x(\{\tilde{x}^a,\hat{x}^a\})\} \nonumber \\
&:=&\ln \prod_a \int \frac{d \tilde{x}^a d \hat{x}^a}{2 \pi} 
\exp \{ \sum_{a \geq b} \hat{q}_{ab}^x \tilde{x}^a \Theta(\tilde{x}^a) \tilde{x}^b \Theta(\tilde{x}^b) 
 +  \kappa \sum_{a,b} R^y_{ab} i \hat{x}^a \tilde{x}^b \Theta(\tilde{x}^b) 
-\frac{1}{2} \sum_{a,b}q^y_{ab} \hat{x}^a \hat{x}^b \nonumber \\
&&-i\sum_{a} \tilde{x}^a \Theta(-\tilde{x}^a)\hat{x}^a -i \nu^x \sum_{a} \hat{x}^a
- \sum_a E_a^x  \tilde{x}^a \Theta(\tilde{x}^a) - \sum_a \hat{p}^x_a \Theta(\tilde{x}^a) \} \nonumber \\
G^y&=&\ln \prod_a \int \frac{d \tilde{y}^a d \hat{y}^a}{2 \pi} 
\exp\{{\cal L}^y(\{\tilde{y}^a,\hat{y}^a\})\} \nonumber \\
&:=&\ln \prod_a \int \frac{d \tilde{y}^a d \hat{y}^a}{2 \pi} 
\exp \{ \sum_{a \geq b} \hat{q}_{ab}^y \tilde{y}^a \Theta(\tilde{y}^a) \tilde{y}^b \Theta(\tilde{y}^b) 
 +  \kappa \sum_{a,b} R^x_{ab} \tilde{y}^a \Theta(\tilde{y}^a) i \hat{y}^b 
-\frac{1}{2} \sum_{a,b}q^x_{ab} \hat{y}^a \hat{y}^b \nonumber \\
&&-i\sum_{a} \tilde{y}^a \Theta(-\tilde{y}^a)\hat{y}^a 
-i \nu^y \sum_{a} \hat{y}^a- \sum_a E_a^y \tilde{y}^a \Theta(\tilde{y}^a) 
- \sum_a \hat{p}^y_a \Theta(\tilde{y}^a) \} \ .
\end{eqnarray}

In the thermodynamic limit $N \to \infty$ the integrals over order parameters in 
(\ref{partrep2}) may be performed by saddle point integration. In order to be able 
to analytically continue the saddle-point to $n \to 0$ we choose the replica-symmetric 
(RS) ansatz for the order parameters 
\begin{eqnarray}
\label{ch3rs}
\begin{array}[h]{lll}
q^{x,y}_{aa}=q^{x,y}_1 & \hat{q}^{x,y}_{aa}=-\frac{1}{2}\hat{q}^{x,y}_1 & \forall a  \\
q^{x,y}_{ab}=q^{x,y}_0 & \hat{q}^{x,y}_{ab}=\hat{q}^{x,y}_0 & \forall a > b \\
R^x_{aa}=R^x_1 & R^y_{aa}=R^y_1 &\forall a \\
R^x_{ab}=R^x_0 & R^y_{ab}=R^y_0 &\forall a \neq b  \\
p^{x,y}_a=p^{x,y} & \hat{p}^{x,y}_a=\hat{p}^{x,y}  &\forall a \\
E^{x,y}_a=E^{x,y} & & \forall a \ .
\end{array}
\end{eqnarray}
$q^{x}_1=\frac{1}{N}\sum_i x_i x_i$ denotes the self-overlap of the mixed strategies of player $X$, 
whereas $q^{x}_0=\frac{1}{N}\sum_i x^1_i x^2_i$ characterizes the overlap between the mixed 
strategies corresponding to two distinct NE and analogously for player $Y$. 

The integrals over $ \tilde{x}^a , \hat{x}^a ,\tilde{y}^a ,\hat{y}^a$ may be evaluated and the 
limit $n \to 0$ may be taken as outlined in appendix \ref{appch31rstext}. 
$G^x$ and $G^y$ evaluated at the RS-saddle point are symmetric with respect 
to an interchange of the players $X$ and $Y$. Thus the maximum of $S(\nu^x,\nu^y)$ occurs at equal  
payoffs and in the thermodynamic limit NE with any other combination of payoffs will be exponentially rare 
by comparison. Hence we may restrict our discussion to the case 
$\nu^x=\nu^y=\nu$, where all order parameters are symmetric under interchange of the players. 

We thus obtain the average entropy of the number of NE within RS 
\begin{eqnarray}
\label{Srep}
S_\kappa(\nu)&=&\frac{1}{N} {\langle \langle} \ln {\cal N}(\nu,\nu) {\rangle \rangle}
=\mbox{2 extr}_{q_1,\hat{q}_1,q_0,\hat{q}_0,R_1,R_0,E,p} 
 \left[\frac{q_1 \hat{q}_1}{2} + \frac{q_0 \hat{q}_0}{2}   - \frac{\kappa R_1^2}{2} + \frac{\kappa R_0^2}{2} 
\right.  \nonumber \\
&& \left. +E -\frac{p}{2} +\int da db \, p_{\tilde{\kappa}}(a,b) \ln L(a,b) \right] \ ,
\end{eqnarray}
where $p_{\tilde{\kappa}}(a,b)$ with $\tilde{\kappa}=\frac{\kappa R_0}{\sqrt{q_0 \hat{q}_0}}$ denotes  
\begin{equation}
\label{ch3pdef}
  p_{\tilde{\kappa}}(a,b) = 
  \frac{1}{2\pi\sqrt{1-\tilde{\kappa}^2}}
  \exp(-\frac{ a^2-2 \tilde{\kappa} a b + b^2 }{2(1-\tilde{\kappa}^2)} ),
\end{equation}
and thus echoes the original 
distribution of the payoffs, and 
\begin{eqnarray}
\label{Srephelp}
L(a,b)&=&H\left(-\frac{\nu-\sqrt{q_0}b}{\sqrt{q_1-q_0}}\right)  \\
&& \hspace{-2.0cm} + \sqrt{\frac{p}{(q_1-q_0)(\hat{q}_1+\hat{q}_0) +\kappa^2 (R_1-R_0)^2}}
G\left(-\frac{\nu-\sqrt{q_0}b}{\sqrt{q_1-q_0}}\right)
K\left( \frac{\frac{\kappa(R_1-R_0)(\sqrt{q_0}b - \nu)}{q_1-q_0} - \sqrt{q_0} a +E} 
{\sqrt{\hat{q}_1+\hat{q}_0 + \frac{\kappa^2(R_1-R_0)^2}{q_1-q_0}}}
\right) , \nonumber 
\end{eqnarray}
where $K(x)$ is a shorthand for $H(x)/G(x)$ with  $G(x)=\frac{1}{\sqrt{2\pi}} \exp(-x^2/2)$, and 
$H(x)=\int_x^{\infty} dy \, G(y)$. The extremum is to be taken over all order parameters, 
The saddle-point equations corresponding to 
(\ref{Srep}) may be solved numerically, their solutions will be discussed in detail in section 
\ref{ch3diss}. 

\subsection{The distribution of the strategy strengths and the potential payoffs}
\label{ch3distr}

In this section we calculate the distribution of strategy strengths 
$\rho_x(x)={\langle \langle} \frac{1}{N} \sum_i \delta(x_i-x) {\rangle \rangle}$ and the potential payoffs 
$\rho_{\lambda_x}(\lambda)={\langle \langle} \frac{1}{N} \sum_i \delta(\sum_j a_{ij} y_j-\lambda) {\rangle \rangle}$ at NE. Due 
to the symmetry of (\ref{partrep2}) under an interchange of players for $\nu^x=\nu^y$ it is sufficient 
to calculate these distributions for a single player only. 
We make use of the set of variables $\tilde{x}_i$ introduced in section 
\ref{ch3indi}, since the distribution $\rho_{\tilde{x}}(\tilde{x})$ is equal to $\rho_x(x)$ for 
$\tilde{x}>0$ and equal to $\rho_{\lambda_x}(\lambda-\nu^x)$ for $\tilde{x}<0$. By the same token, the 
fraction of strategies with $x_i=0$ is equal to 
$\int_{-\infty}^0 d \tilde{x} \rho_{\tilde{x}}(\tilde{x})=1-p$ and the fraction of potential payoffs with 
$\lambda^x_i=\nu^x$ is $\int_{0}^{\infty} d \tilde{x} \rho_{\tilde{x}}(\tilde{x})=p$. 

Since all pure strategies are equivalent after averaging over the payoffs (translation-invariance), 
we calculate 
\begin{eqnarray}
\label{appdisteq}
\rho_{\tilde{x}}(\tilde{x}) &=& \left\langle \left\langle   \frac
{\int d \mu({\bf \tilde{x}}) d \mu ({\bf \tilde{y}}) \delta(\tilde{x}_1-\tilde{x})
   \prod_i \delta(I_i^x) \prod_j \delta(I_j^y) 
   \| \frac {\partial ({\bf I}^x,{\bf I}^y)}{\partial ({\bf \tilde{x}},{\bf \tilde{y}})} \| }
{\int d \mu({\bf \tilde{x}},{\bf \tilde{y}}) 
   \prod_i \delta(I_i^x) \prod_j \delta(I_j^y) 
   \| \frac {\partial ({\bf I}^x,{\bf I}^y)}{\partial ({\bf \tilde{x}},{\bf \tilde{y}})} \| } \right\rangle \right\rangle  \nonumber \\
 &=& \lim_{n \to 0} {\langle \langle} \prod_{a=1}^n \int d \mu({\bf \tilde{x}},{\bf \tilde{y}}) \delta(\tilde{x}^1_1-\tilde{x})
   \prod_{i,a} \delta(I_i^{xa}) \prod_{j,a} \delta(I_j^{ya}) 
   \prod_a \| \frac {\partial ({\bf I}^{xa},{\bf I}^{ya})}{\partial ({\bf \tilde{x}}^a,{\bf \tilde{y}}^a)} \|
 {\rangle \rangle}
\ .
\end{eqnarray}
In order to be able to perform the average over payoffs occurring in both the numerator and the 
denominator, we have represented the denominator by $n-1$ replicas. The average over payoffs now 
proceeds exactly as in the previous section. Introducing the matrices of order parameters $q_{ab}$ 
again, the $i=1$ term may be split off from the saddle point integral without distorting the saddle 
point in the thermodynamic limit. Taking the replica symmetric ansatz one obtains 
\begin{eqnarray}
\rho_{\tilde{x}}(\tilde{x})&=& \lim_{n \to 0} \prod_a \int \frac{d \tilde{x}^a d \hat{x}^a}{2\pi} \exp\left\{{\cal L}^x(\{\tilde{x}^a,\hat{x}^a\})\right\} 
\delta(\tilde{x}^1-\tilde{x}) \\
&=& \left\{ \begin{array}{ll}  
\int da db \,  p_{\tilde{\kappa}}(a,b) {\frac{1}{\sqrt{2\pi(q_1-q_0)}{L(a,b)}}\exp\{-\frac{(\tilde{x}+\nu-\sqrt{q_0}b)^2}{2(q_1-q_0)}\}}
&  \ \ \tilde{x} < 0 \\
& \\  
\int da db\,  p_{\tilde{\kappa}}(a,b) \frac{\sqrt{p}}{\sqrt{2\pi(q_1-q_0)}{L(a,b)}} \\
\hspace{2.5cm}
\exp\{-\frac{(-\kappa(R_1-R_0)\tilde{x}+\nu-\sqrt{q_0}b)^2}{2(q_1-q_0)} 
-\frac{1}{2}(\hat{q}_1+\hat{q}_0)\tilde{x}^2+a\sqrt{\hat{q}_0}\tilde{x}-E\tilde{x}\}   
&  \ \ \tilde{x} > 0 ,   
\end{array} \right.  \nonumber 
\end{eqnarray}
where ${\cal L}^x(\{\tilde{x}^a,\hat{x}^a\})$ was defined in (\ref{Gdef}), the order parameters take on 
their saddle point values, and we have dropped the 
player-indices of the order parameters. These functions have to be evaluated numerically 
and will be discussed in section \ref{ch3dissstab}.  

The same procedure may be used to calculate another quantity of interest, namely the 
fraction $w$ of pure strategies which are \emph{both} played with non-zero probability in two 
randomly chosen NE equilibria. Like $q_0$, this quantity is a measure of the 
degree of similarity of two randomly chosen NE. However $w$ does not directly depend on 
the self-overlap of the mixed strategies and may serve to test if there are strategies which are 
more likely to be played at a NE than others. 
From (\ref{partrep2}) and (\ref{ch3rs}) one obtains
\begin{eqnarray}
w&=& \lim_{n \to 0} \prod_a \int \frac{d \tilde{x}^a d \hat{x}^a}{2\pi} \exp\left\{{\cal L}^x(\{\tilde{x}^a,\hat{x}^a\})
\right\} 
\Theta(\tilde{x}^1)\Theta(\tilde{x}^2) \\
&=& 
\int da db\,  p_{\tilde{\kappa}}(a,b) 
\frac{p}{\left((q_1-q_0)(\hat{q}_1+\hat{q}_0) +\kappa^2 (R_1-R_0)^2\right){L^2(a,b)}} \nonumber \\
&& \hspace{2.6cm}
G^2\left(-\frac{\nu-\sqrt{q_0}b}{\sqrt{q_1-q_0}}\right)
K^2\left( \frac{\frac{\kappa(R_1-R_0)(\sqrt{q_0}b - \nu)}{q_1-q_0} - \sqrt{q_0} a +E} 
{\sqrt{\hat{q}_1+\hat{q}_0 + \frac{\kappa^2(R_1-R_0)^2}{q_1-q_0}}}\right)
 \nonumber \ .
\end{eqnarray}

\subsection{The stability of the replica-symmetric saddle point}
\label{ch3rsstab}

The results for the quenched average were derived on the basis of the
replica-symmetric ansatz (\ref{ch3rs}).  In this section we
investigate the stability of this ansatz with respect to small
fluctuations around (\ref{ch3rs}) in order to check if this ansatz is
at least locally stable. We restrict ourselves to the
special case $\kappa=0$, where the payoff matrices are uncorrelated
and the order parameters $R_{ab}^x,R_{ab}^y$ do not arise \cite{probskappa}. We consider 
small transversal fluctuations around
the RS saddle-point and expand (\ref{partrep2}) to second order in
these fluctuations to obtain
\begin{equation}
 S=S_{RS} + \frac{1}{2} \Delta^T M \Delta + {\cal O}(\Delta^3) \ ,
\end{equation} 
where $\Delta$ denotes a vector of small fluctuations in the
off-diagonal elements of the order parameters
$q_{ab}^x,\hat{q}^x_{ab},q_{ab}^y,\hat{q}^y_{ab}$ and $M$ is given by 
\begin{equation}
\label{ch3Mdef}
  M=\left(
  \begin{array}{llll}
  \frac{\partial^2 G^y}{\partial q^x_{ab} \partial q^x_{cd}} & -I & 0 & 
  \frac{\partial^2 G^y}{\partial q^x_{ab} \partial \hat{q}^y_{cd}}   \\
  -I &\frac{\partial^2 G^x}{\partial \hat{q}^x_{ab} \partial \hat{q}^x_{cd}} & 
  \frac{\partial^2 G^x}{\partial \hat{q}^x_{ab} \partial q^y_{cd}} &0\\
  0 &\frac{\partial^2 G^x}{\partial q^y_{ab} \partial \hat{q}^x_{cd}}& 
  \frac{\partial^2 G^x}{\partial q^y_{ab} \partial q^y_{cd}} & -I \\
 \frac{\partial^2 G^y}{\partial \hat{q}^y_{ab} \partial q^x_{cd}} & 0 &
 -I& \frac{\partial^2 G^y}{\partial \hat{q}^y_{ab} \partial \hat{q}^y_{cd}}
  \end{array} \right) \ .
\end{equation}
Due to the symmetry of (\ref{partrep2}) under an interchange of the players at the RS saddle-point, 
only 3 different non-trivial submatrices need to be evaluated. 
The criterion for RS to be locally stable needs to be determined by working out the 
paths of integration in the complex plane of $\hat{q}^x$ and $\hat{q}^y$. 
Denoting the replicon-eigenvalues of $\frac{\partial^2 G^x}{\partial \hat{q}^x_{ab} \partial \hat{q}^x_{cd}}$, 
$\frac{\partial^2 G^y}{\partial q^x_{ab} \partial q^x_{cd}}$, and 
$\frac{\partial^2 G^y}{\partial q^x_{ab} \partial \hat{q}^y_{cd}}$ by $\lambda_1,\lambda_2,\lambda_3$ respectively 
one obtains the criterion for the local stability of RS
\begin{eqnarray}
\frac{1}{\lambda_2}(\lambda_1 \lambda_2 -(\lambda_3-1)^2) &<& 0 \nonumber \\
\frac{1}{\lambda_2}(\lambda_1 \lambda_2 -(\lambda_3+1)^2) &<& 0	\ .
\end{eqnarray}
For details of the calculation see appendix \ref{appch31rsstab}. 

\section{Discussion of the results}
\label{ch3diss}

The quantity $S_\kappa(\nu)$ contains a wealth of information. We begin by discussing 
the general shape of $S_\kappa(\nu)$ and the number of NE as a function of $\kappa$, then turn to 
the statistical properties of NE and give a geometric interpretation of the results, and finally 
discuss the distribution of potential payoffs and the strategy strengths. 

\subsection{$S_\kappa(\nu)$ and the number of NE}

\begin{figure}[htb]
	\epsfysize=7.5cm
       \epsffile{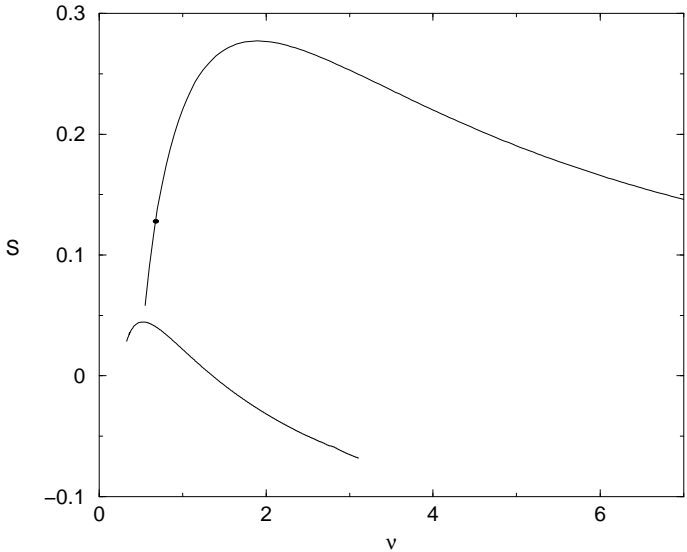}
\caption{\label{figkappa0} The results of the quenched averages of 
$S_{\kappa}(\nu)$ for $\kappa=-0.8$ and $\kappa=0$ respectively (bottom and top respectively). 
For $\kappa=0$ replica symmetry is locally stable for $\nu>0.67$ as indicated by the black dot.}
\end{figure}

The expression (\ref{Srep}) for  
$S^{RS}_{\kappa}(\nu)$ defines a family of curves with a pronounced maximum 
shown exemplarily for $\kappa=0$ in figure \ref{figkappa0}. 
As argued in section \ref{ch3indi} in the thermodynamic limit the maximum of $S_{\kappa}(\nu)$  
dominates the spectrum of NE. 

Another point of interest is the value of $\nu$ where $S_{\kappa}(\nu)$ crosses the $S=0$ axis. Due to the symmetry of 
$S(\nu^x,\nu^y)$ under an interchange of the players this point indicates the 
NE with the maximum sum of the payoffs. For $\kappa=-1$ it takes on the value $0$ and increases monotonously 
with $\kappa$. At $\kappa=\kappa_c$ it diverges to infinity; $S_{\kappa}(\nu)$ no longer crosses the $S=0$ 
axis. In this case there is an exponentially large number of NE offering an 
arbitrarily large payoff to either player, where an arbitrarily small fraction of strategies are 
played. From the annealed approximation one obtains $\kappa_c \approx -.59$, the corresponding 
result from the RS expression for the quenched average is $\kappa_c \approx -.58$. This effect may be 
explained as follows: At large values of $\kappa$ players may pick a few of the pairs of strategies 
$(i,j)$, which offer a large payoff to both of them and play them with a large probability. 
An exponential number of NE may be constructed in this way, even though their number is 
exponentially small compared to the total number of NE. 

\begin{figure}[htb]
	\epsfysize=7.5cm	
       \epsffile{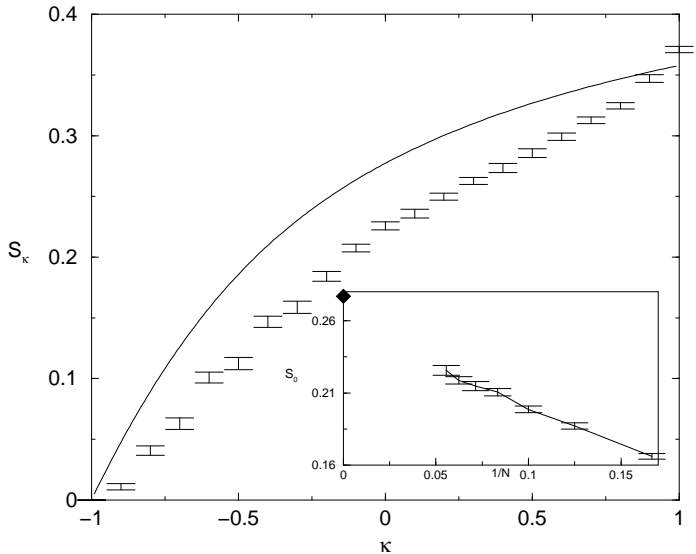}
\caption{\label{figSk}The RS entropy of NE $S_{\kappa}$ as a function of $\kappa$  (solid line). 
The numerical results stem from enumerations with system size N=18 averaged over 100 samples, the 
error-bars denote the statistical error.  
The inset shows the finite-size effects for the case $\kappa=0$. $S_0$ is plotted against $1/N$, 
the analytic result for $N \to \infty$ is indicated by the filled diamond.}
\end{figure}

The entropy of NE given by the maximum $S_{\kappa}$ of $S_{\kappa}(\nu)$ is shown in figure 
\ref{figSk}. 
We find $S_{\kappa=-1}=0$ since there is only a single equilibrium 
point for zero-sum games. $S_{\kappa}$ increases with $\kappa$, so for all $\kappa>-1$ the typical 
number of NE scales exponentially with $N$. The maximum of the typical number of NE is reached for 
the case of symmetric games, where $S_{\kappa=1} \sim 0.358$. This result may be compared with a 
rigorous upper bound for the maximal number of NE in a bimatrix game derived using geometric 
methods \cite{Keiding,Stengel}: For any non-degenerate 
$N$-by-$N$ bimatrix game with large $N$ there are at most $e^{0.955 N}$ equilibrium points. 
As expected, the typical-case scenario does not saturate this bound, at least not for the distribution 
of payoffs considered here. Nevertheless for $\kappa>-1$ the typical number of NE investigated here 
and the maximal number of NE both scale exponentially with $N$.  

The increase of the number of NE with the correlation between the payoff matrices may be explained 
as follows: As will be discussed in the next section, the payoff $\nu$ to both players increases 
with $\kappa$. For increasing values of $\nu=\nu^x=\nu^y$ the necessary (but not sufficient) 
conditions for a NE 
\begin{eqnarray}
\label{NEnecc}
  \sum_j a_{ij} y_j  \leq  \nu^x & \, x_i \geq 0  & \forall i \nonumber \\
 \sum_i x_i b_{ij}   \leq  \nu^y & \, y_j \geq 0  & \forall j 
\end{eqnarray}
become increasingly easy to fulfill. In fact for $\nu=0$ only a single point on the simplexes of the 
two players fulfills (\ref{NEnecc}), whereas for large $\nu$ a correspondingly large 
section of the simplexes qualify as a candidate for equilibrium points \cite{BergEngel}. 
As a result the number of points which apart from (\ref{NEnecc}) obey $(\sum_j a_{ij} y_j-\nu^x) x_i =0$ 
and $( \sum_i x_i b_{ij}- \nu^y)y_j=0$ and thus constitute NE increases with $\kappa$. 

\subsection{The statistical properties of Nash equilibria}

In the thermodynamic limit not only the number of NE will be dominated by the maximum of 
$S_{\kappa}(\nu)$, but a randomly chosen NE will also give the payoff
$\nu=\nu^x=\nu^y=\mbox{argmax}S_\kappa(\nu)$ with probability one, because the number of 
NE with this payoff is exponentially larger than the number of all other NE. Similarly, the selfoverlap,
the mutual overlap, and the fraction of strategies played
with non-zero probability will take on their saddle-point
values evaluated at the maximum of $S_{\kappa}(\nu)$ 
with probability 1. 

\begin{figure}[htb]
 \epsfysize=7.5cm
      \epsffile{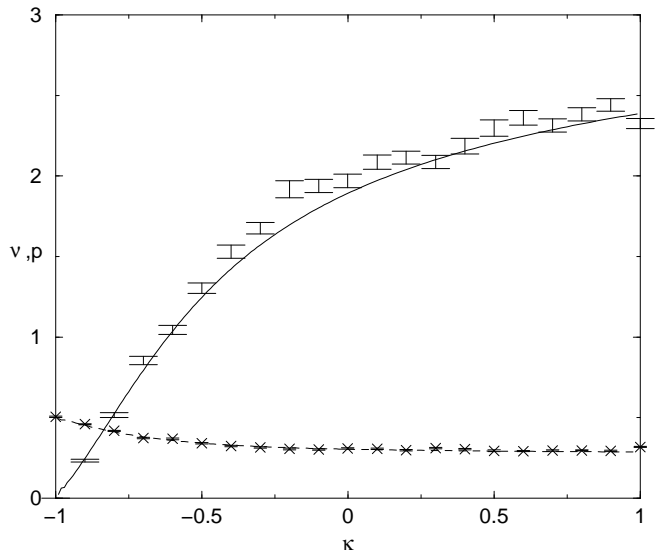}
\caption{
\label{fignup}
The payoff $\nu$ (solid line) and the fraction $p$ (dashed line) of strategies played with non-zero 
probability of the typical NE. The analytic results are compared with numerical simulations for 
$N=50$ averaged over 200 samples.
}
\end{figure}

\begin{figure}[htb]
 \epsfysize=7.5cm
      \epsffile{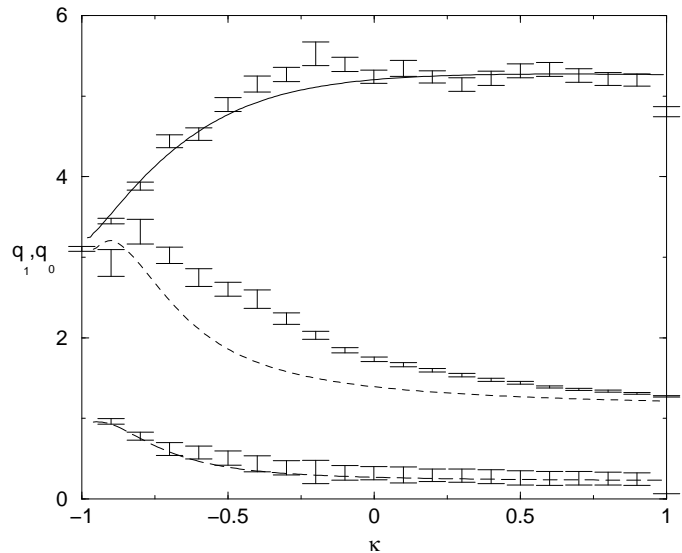}
\caption{
\label{figq01}
The self-overlap of mixed strategies $q_1$ (solid line), the overlap $q_0$  between 
the mixed strategies of different NE (dashed line), and the 
ratio $q_0/q_1$ (long-dashed line).  
The analytic results are compared with numerical simulations for $N=50$ averaged
over 200 samples for $q_1$  and $N=18$ averaged over 100 samples for $q_0$.
}
\end{figure}

Figure \ref{fignup} shows the payoff dominating the spectrum of NE and the corresponding 
fraction $p$ of strategies played with non-zero probability, whereas the self-overlap of mixed 
strategies $q_1$ and the overlap $q_0$ between the mixed strategies of different NE are shown 
in figure \ref{figq01}. 

At $\kappa=-1$ we recover the results for zero-sum games with $q_1=q_0=\pi$, $\nu=0$, 
and $p=1/2$ \cite{BergEngel}. As $\kappa$ rises, the payoff increases. This effect may 
be understood as follows: At increasing $\kappa$ the outcome of a pair of strategies
$(i,j)$ which is beneficial to player $X$ say, tends to become more
beneficial to player $Y$.  As a result players focus on these
strategies and the payoff at a NE to both players rises.  By the same token, 
the fraction $p$ of strategies which are played with non-zero
probability at a NE decreases with $\kappa$ and the self-overlap $q_1$ of
the mixed strategies increases.

The geometric structure of the set of NE may be elucidated by considering the mutual overlap 
$q_0=\frac{1}{N} \sum_i x^1_i x^2_i$ 
between mixed strategies of different NE. At $\kappa=-1$, where there is only a single NE, 
$q_0$ equals the self-overlap  
$q_1$. After an initial increase $q_0$ decreases with increasing $\kappa$.  
The initial increase of $q_0$ is due to the rapid increase of the lengths of the 
mixed strategy vectors and is thus not seen in the ratio between the overlaps.   
This result may be interpreted geometrically in that  
the NE become more and more separated with increasing $\kappa$, 
and for $\kappa \to +1$ they end up in nearly uncorrelated positions, 
$\langle x^1x^2\rangle-\langle x^1\rangle\langle x^2\rangle=.21$.  
At the same time an increasing fraction of components of the mixed strategies have $x_i=0$, 
i.e. lie on an edge of the simplex. 

\begin{figure}[htb]
 \epsfysize=7.5cm
      \epsffile{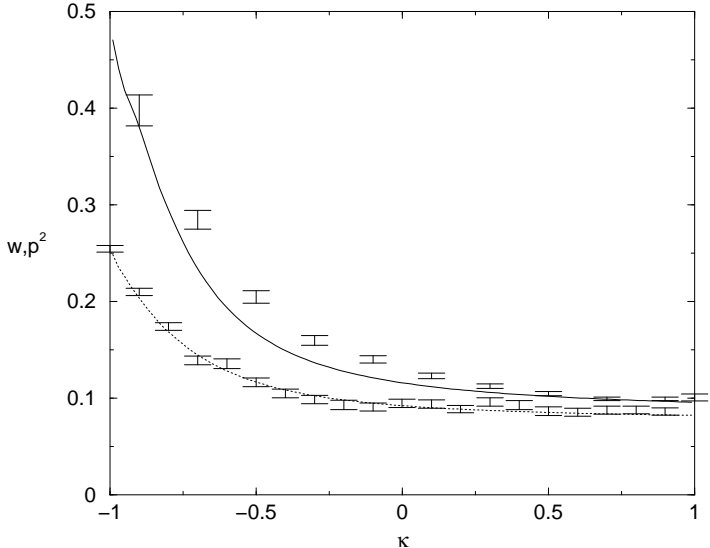}
\caption{
\label{figbp}
The fraction of strategies $w$ (top) played in both mixed strategies of two randomly chosen NE 
and the square $p^2$ of the fraction of strategies played at a single NE (bottom) against $\kappa$. 
The analytic results are compared with numerical simulations for $N=50$ averaged
over 200 samples for $p^2$  and $N=18$ averaged over 100 samples for $w$.}
\end{figure}

Even though the NE spread over the simplex with increasing $\kappa$, players still tend to focus 
on specific strategies. This may be seen by comparing the 
fraction of strategies $w$ played in \emph{both} mixed strategies of two randomly chosen NE with 
the corresponding result $p^2$, which would result if 
players chose $p$ strategies to be played with non-zero probabilities at random. From figure \ref{figbp} 
one finds that although $w$ decreases with $\kappa$ consistent with the spread of NE over the simplex 
it always remains above $p^2$. This behaviour is consistent with the idea that with increasing 
$\kappa$ players focus on pairs of strategies which are beneficial to both, of which there is a 
large number for large values of $\kappa$. 

Since NE are isolated points, replica symmetry describes a set of equilibrium points 
which are distributed uniformly over a part of the simplex with opening angle $\arccos (q_0/q_1)$. 
A replica-symmetry breaking scenario would involve clusters of NE, and maybe even clusters of 
clusters, so an ansatz explicitly including more than two overlap scales would have to be employed 
along the lines of the Parisi-scheme \cite{MPV}. 
However, at least for $\kappa=0$, we found that replica symmetry is locally stable 
for $\nu>0.67$ and most importantly at the maximum of the curve. RS remained locally stable across the 
range of $\nu$ investigated, nevertheless RS may become locally unstable again at sufficiently large 
values of $\nu$. Thus we may 
conclude that for typical NE at $\kappa=0$ the replica-symmetric ansatz is self-consistent. 
Since we know from the results of \cite{BergEngel} that RS is marginally stable at $\kappa=-1$ 
one may in fact speculate that for the typical NE the RS scenario holds across the 
entire range of $\kappa$. Nevertheless there may well be distributions of the payoffs which 
lead to non-uniform distributions of NE and to replica-symmetry breaking, presumably distributions 
with large values of $\kappa$, or with correlations between the entries of the payoff 
matrices at different sites. 

\subsection{The distribution of potential payoffs and the strategy strengths}
\label{ch3dissstab}

\begin{figure}[h]
\begin{minipage}[t]{0.55 \textwidth}
 \epsfxsize= 0.8\textwidth
\epsffile{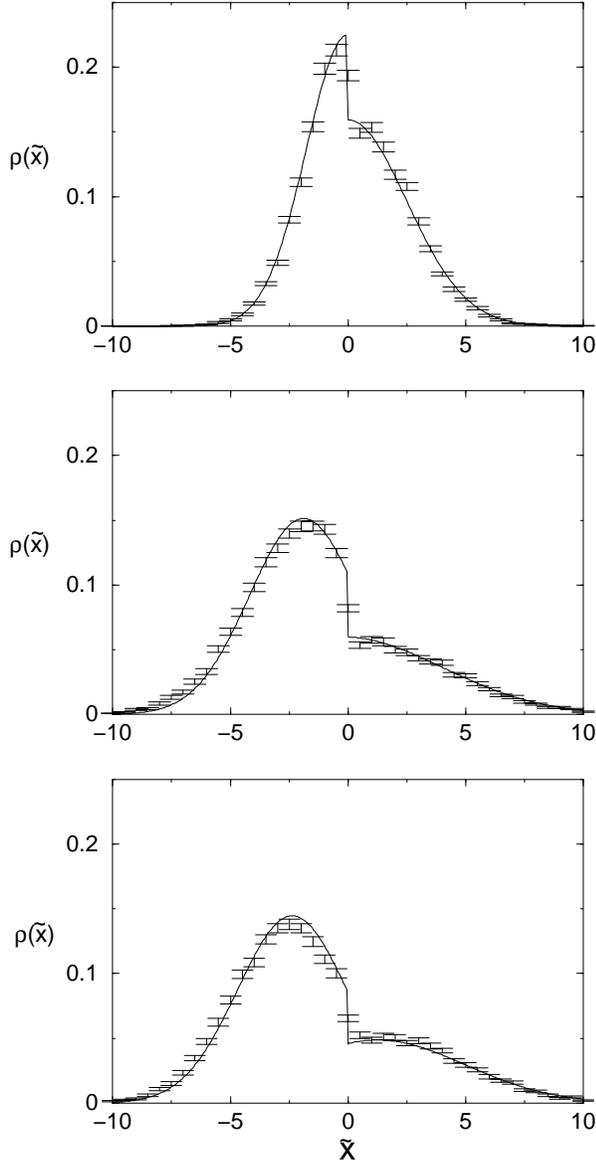}
\end{minipage}
\begin{minipage}[t]{0.42 \textwidth}
\vspace{-6.5cm}  
\caption{The distribution of the potential payoffs ($\tilde{x}<0$) and the strategy strengths 
($\tilde{x}>0$) for $\kappa=-1,0,1$ from top to bottom}
\label{figstabs}
\end{minipage}
\end{figure}

Figure \ref{figstabs} shows the distribution of strategy strengths 
$\rho_x(x)={\langle \langle} \frac{1}{N} \sum_i \delta(x_i-x) {\rangle \rangle}$ ($\tilde{x}>0$) and the potential payoffs 
$\rho_{\lambda_x}(\lambda)={\langle \langle} \frac{1}{N} \sum_i \delta(\sum_j a_{ij} y_j-\lambda) {\rangle \rangle}$ 
($\tilde{x}<0$) 
calculated in section \ref{ch3distr}. The decrease of the fraction of strategies played with non-zero 
probability, $\int_{0}^{\infty} d \tilde{x} \rho_{\tilde{x}}(\tilde{x})=p$ with $\kappa$ is clearly 
visible. One also finds a marked tendency for both players to use large values of $x_i$ and $y_j$ for decreasing 
values of $p$, as is demanded by the normalization condition. 

One also observes the formation of a `shoulder' in the distribution of 
$\rho_{\tilde{x}}(\tilde{x})=\rho_{\lambda_x}(\lambda-\nu^x)$
($\tilde{x}<0$) centered at $-\nu$. It shows that the distribution of $\sum_j a_{ij} y_j$ remains 
peaked at zero leading to the formation of the shoulder at $\tilde{x}=-\nu$ as $\nu$ increases. 

\subsection{Comparison with numerical results}

The numerical results for figure \ref{figSk}, \ref{figstabs}, $q_0$ in \ref{figq01}, and 
$w$ in \ref{figbp} were obtained 
by using so-called vertex enumeration to enumerate all NE. Since the computational effort for vertex 
enumeration scales with $\sim 2.598^{N}$, the system size had to be restricted to $N=18$ and 
averages were taken over 100 samples, resulting in pronounced finite size effects. Nevertheless 
the increase of the number of NE with $\kappa$ is clearly confirmed by the simulations.  

The numerical results for figure \ref{fignup}, $q_1$ in figure \ref{figq01}, $p$ in \ref{figbp}, 
and figure \ref{figstabs}   
were obtained by using an iterated variant of the Lemke-Howson algorithm \cite{LH,Stengel} to 
locate \emph{a single} NE and 
by averaging the results for $N=50$ over 200 different realizations of the payoffs. 
Although some finite-size effects remain, there is good agreement between the analytical and 
the numerical results. 

\section{Summary and outlook}

We analysed the properties of Nash equilibria in large random bimatrix games.  
To this end we constructed an indicator function which was used to count the number of NE 
with given payoffs to both players. 
We found that the number of NE is exponentially dominated by NE with a certain payoff 
to both players, and a certain set of order parameters. 
This implies that    
for a randomly chosen Nash equilibrium quantities such as the fraction 
of strategies played with a given probability, the self-overlap,  
and most importantly the payoff to either player take on a specific
value with probability 1. 

We considered square payoff matrices and argued that for large games and identically and 
independently distributed elements of the payoff matrices at different sites $(i,j)$, 
the only relevant parameter of the probability ensemble is the 
correlation between elements of the same site of the payoff matrices $a$ and $b$. We then calculated the 
quenched average of the number of Nash equilibria, from which one may also deduce 
quantities such as the payoff, the self-overlap and the mutual overlap of mixed strategies at NE, and 
the distribution of the strategy strengths and the potential payoffs.    

We found that both the number of equilibrium points and the payoff to both players increase with the 
correlation between the payoff matrices: With increasing correlation the number of pairs of strategies 
which are beneficial to both players grows. Players may focus on these pairs and achieve a larger 
payoff, the 
fraction of strategies played with non-zero probability decreases accordingly. From the values of the 
saddle-point parameters one may also deduce information on the geometric properties of NE: With 
increasing correlation between the payoff matrices the NE spread out over wider regions of the simplex. 
These analytic results were quantitatively compared with extensive numerical simulations and  
good agreement was found. 

Another point of relevance is that for a sufficiently large correlation between the payoff matrices, 
an exponentially large number of NE appears which offer arbitrarily large payoffs (on the statistical 
mechanics scale) to both players. The number of such NE is of course exponentially small compared 
to the total number of NE, nevertheless these equilibrium points may be relevant if players are 
free to choose equilibrium points. 

A number of generalizations and extensions of these scenarios may be envisaged at this stage, 
including the investigation of bimatrix games with rectangular payoff matrices or payoff matrices 
with correlations between the elements at different sites. 
Furthermore, a scenario of games of several players might be extended to describe cooperative games, 
where coalitions of players pool their payoffs and seek to maximize the 
gain of their respective coalition. In this context it may also be interesting to consider the 
case of ${\cal O}(N)$ players choosing between ${\cal O}(1)$ strategies.

{\bf Acknowledgments}: I would like to thank A.~Engel, M.~Opper, and M.~Weigt for fruitful discussions 
and the Studien\-stiftung des Deutschen Volkes for financial support. 

\appendix

\section{The quenched average}
\label{appendix}

In the following we give a derivation of the quenched average of the entropy of NE. 
In order to represent the logarithm of ${\cal N}$ we replicate $n$ times the expression for the number 
of NE ${\cal N}$ (\ref{ch3part}) to obtain (\ref{partrep1}). 

Treating the normalizing determinant as a self-averaging quantity, we may split off $\ln \|\det(D)\|$ with
\begin{equation}
\label{appch31Bdef}	
  D=
  \left(
  \begin{array}{ll}
  \delta_{ii'}\Theta(-\tilde{x_i}) & -a_{ij}\Theta(\tilde{y_j}) \\
   -b_{ij}\Theta(\tilde{x_i}) & \delta_{jj'}\Theta(-\tilde{y_j})  \\
  \end{array} \right) \ .
\end{equation}  
from (\ref{partrep1}) and \emph{separately} average the normalizing determinant over the disorder.  
Leaving out all the rows and columns which have only the entry $1$ along the diagonal 
and do not contribute to the determinant, we are left with the determinant of a matrix 
\begin{equation}
\label{appch31Bprime}
D'= \left(
  \begin{array}{ll}
  0 & -a' \\
   -b'& 0
    \end{array} \right) \ ,
\end{equation}  
where the matrices $a'$ and $b'$ are the $pN$ by $pN$ submatrices of the payoff matrices 
containing the elements with $\tilde{x}_i>0$ and $\tilde{y}_j>0$. 
We thus calculate ${\langle \langle} \| \ln \det(D) \| {\rangle \rangle}$ as a function of $p_x=p_y$ exploiting the 
block-structure of the matrix $D$ and using results from the theory of random matrices \cite{CPV}. 
Since we have $\ln \| \det(D)\| =\ln \| \det(a')\|+\ln \| \det(b') \|$, 
the correlation between the elements of these matrices has no effect. We may thus use the 
circular theorem \cite{Wigner}, which gives the average density 
$\rho(\omega)$ of eigenvalues $\omega$ of a $pN$ by $pN$ matrix with identically and independently 
Gaussian distributed entries with zero mean and variance $N^{-1}$ 
\begin{equation}
\rho(\omega)=\left\{
\begin{array}{ll}
(\pi p)^{-1} & \| \omega \|<\sqrt{p} \\
0 & \mbox{otherwise}
\end{array}
\right.
\end{equation}
giving 
\begin{equation}
\label{appch31qudet}
	{\langle \langle} \ln \| \det(D) \| {\rangle \rangle}=2Np \int_S d\omega \rho(\omega) \| \omega \| = Np(\ln p-1) \ ,
\end{equation}
where the integral is over the region in the complex plane with $\| \omega \|<\sqrt{p}$. 
After this step, the only terms in (\ref{partrep1}) where the disorder is present are 
\begin{eqnarray}
\label{appch31av}
{\langle \langle} \prod_{i,j} && \exp \{ i \sum_a \hat{x}^a_i a_{ij} \tilde{y}^a_j \Theta(\tilde{y}^a_j) + 
                         i \sum_a\tilde{x}^a_i \Theta(\tilde{x}^a_i) b_{ij} \hat{y}^a_j  \}{\rangle \rangle} = \\
&&\exp\{-1/(2N)\sum_{a,b} (\sum_i \hat{x}^a_i \hat{x}^b_i \sum_j  \tilde{y}^a_j \Theta(\tilde{y}^a_j) 
\tilde{y}^b_j \Theta(\tilde{y}^b_j) \nonumber \\
&&              -2 \kappa \sum_i i \hat{x}^a_i \tilde{x}^b_i \Theta(\tilde{x}^b_i) 
                             \sum_j  i\hat{y}^b_j \tilde{y}^a_j \Theta(\tilde{y}^a_j) 
            +  \sum_i  \tilde{x}^a_i\Theta(\tilde{x}^a_i) \tilde{x}^b_i \Theta(\tilde{x}^b_i) 
	\sum_j \hat{y}^a_j \hat{y}^b_j ) \} \ , \nonumber
\end{eqnarray}
where the indices $a$ and $b$ denote the replicas, $a,b=1 \dots n$ and the average has been taken over the 
distribution of payoffs (\ref{eq:payensemble}). In order to obtain expressions which factorize in $i$ and 
$j$ we introduce the matrices of order parameters  
\begin{eqnarray}
\label{ch3ops}
q_{ab}^x=\frac{1}{N} \sum_{i} \tilde{x}_i^a \Theta(\tilde{x}_i^a) \tilde{x}_i^b \Theta(\tilde{x}_i^b)  \ , \ 
&&q_{ab}^y=\frac{1}{N} \sum_{j} \tilde{y}_j^a \Theta(\tilde{y}_j^a) \tilde{y}_j^b \Theta(\tilde{y}_j^b) \nonumber \\
R_{ab}^x=\frac{1}{N} \sum_{i} i \hat{x}^a_i \tilde{x}_i^b \Theta(\tilde{x}_i^b)   \ , \ 
&&R_{ab}^y=\frac{1}{N} \sum_{j} i \hat{y}^b_j \tilde{y}_j^a \Theta(\tilde{y}_j^a) \nonumber \\
p_{a}^x=\frac{1}{N} \sum_{i} \Theta(\tilde{x}_i^a)    \ , \ 
&&p_{a}^y=\frac{1}{N} \sum_{j} \Theta(\tilde{y}_j^a) \ ,
\end{eqnarray} 
using integrals over delta functions. The last pair of order parameters is introduced so the normalizing 
determinant may be included as a function of $p_a^x$ and $p_a^y$. This procedure turns (\ref{appch31av}) 
into 
\begin{eqnarray}
\label{appch31av2}
 && \prod_{a \geq b}\int \frac{dq_{ab}^{x,y} d\hat{q}_{ab}^{x,y}}{2 \pi/N} 
  \prod_{a,b}\int \frac{dR_{ab}^x dR_{ab}^y} {2 \pi/(\kappa N)}
  \prod_a \int \frac{dp_{a}^{x,y} d\hat{p}_{a}^{x,y}}{2 \pi/N} \, \delta(p_a^x-p_a^y)  \nonumber \\
&& \exp \{ -iN \sum_{a \geq b} q_{ab}^{x,y} \hat{q}^{x,y}_{ab} -i \kappa N \sum_{a,b} R_{ab}^x R_{ab}^y + 
  iN \sum_a p^{x,y}_a \hat{p}^{x,y}_a   \} \nonumber \\
&&\exp \{ \sum_{a \geq b} \hat{q}_{ab}^x \tilde{x}^a \Theta(\tilde{x}^a) \tilde{x}^b \Theta(\tilde{x}^b) 
 + i \kappa \sum_{a,b} R^y_{ab} i \hat{x}^a \tilde{x}^b \Theta(\tilde{x}^b) -\frac{1}{2} \sum_{a,b}q^y_{ab} \hat{x}^a \hat{x}^b \nonumber \\
&&-i\sum_{a,i} \tilde{x}^a_i \Theta(-\tilde{x}^a_i)\hat{x}^a_i -i \nu^x \sum_{a,i} \hat{x}^a_i -i \sum_a \hat{p}^x_a \Theta(\tilde{x}^a) \} \nonumber \\
&& \exp \{ \sum_{a \geq b} \hat{q}_{ab}^y \tilde{y}^a \Theta(\tilde{y}^a) \tilde{y}^b \Theta(\tilde{y}^b) 
 +  \kappa \sum_{a,b} R^x_{ab} \tilde{y}^a \Theta(\tilde{y}^a) i \hat{y}^b -\frac{1}{2} \sum_{a,b}q^x_{ab} \hat{y}^a \hat{y}^b \nonumber \\
&&-i\sum_{a,i} \tilde{y}^a_i \Theta(-\tilde{y}^a_i)\hat{y}^a_i -i \nu^y \sum_{a,i} \hat{y}^a_i -i \sum_a \hat{p}^y_a \Theta(\tilde{y}^a) \}
\end{eqnarray}
All order parameters have been introduced via conjugate variables, except $R_{ab}^x$ and $R_{ab}^y$, which are 
conjugate to each other. Care must be taken to scale all order parameters so they are of ${\cal O}(1)$ in 
the thermodynamic limit. Expression (\ref{appch31av2}) may now be substituted back into (\ref{partrep1}). 
The simplex-constraint 
is incorporated by including yet another set of integrals 
\begin{equation}
\prod_a \int \frac{dE_a^{x,y}}{2\pi/N} \exp \{i N\sum_a E^{x,y}_a 
-i \sum_a E_a^x \sum_i \tilde{x}^a_i \Theta(\tilde{x}^a_i)-i \sum_a E_a^y \sum_i \tilde{y}^a_i \Theta(\tilde{y}^a_i) \} \ .
\end{equation} 
The integrals over $\tilde{x}_i^a, \hat{x}_i^a$ now factorize and form a product of $N$ identical terms and may 
thus be written as the $N$-th power of a single such term. The same point applies to the integrals over 
$\tilde{y}_j^a, \hat{y}_j^a$. Anticipating saddle points of the integrals over conjugate order parameters 
along the imaginary axis, we also perform a change of variables 
$i\hat{q}^{x,y}_{ab} \rightarrow \hat{q}^{x,y}_{ab}$ 
and analogously for $R^y_{ab},E^{x,y}_a,\hat{p}^{x,y}_a$. Including the normalizing determinant 
(\ref{appch31qudet}) we finally obtain (\ref{partrep2}) and (\ref{Gdef}). 

\subsection{The replica symmetric ansatz}  
\label{appch31rstext}

In the thermodynamic limit $N \to \infty$ the integrals over order parameters are dominated by their 
saddle point. Yet in order to carry out the replica limes $n \to 0$ we have to make an ansatz 
for the values of the order parameter matrices. The simplest ansatz is the replica-symmetric 
one given by (\ref{ch3rs}). 
Since $G^x$ and $G^y$ are symmetric under interchange of the players we may drop the 
superscripts $x$ and $y$. We obtain 
\begin{eqnarray}
G&=&
\ln \prod_a \int \frac{d \tilde{x}^a d \hat{x}^a}{2 \pi} \exp \{ -\frac{1}{2}\sum_{a} (\hat{q}_1+\hat{q}_0) \tilde{x}^a \tilde{x}^a  \Theta(\tilde{x}^a) 
+\frac{1}{2}\sum_{a,b} \hat{q}_0 \tilde{x}^a \Theta(\tilde{x}^a) \tilde{x}^b \Theta(\tilde{x}^b)  \\	 	
&& +  \kappa (R_1-R_0) \sum_{a} i \hat{x}^a \tilde{x}^a \Theta(\tilde{x}^a) +\kappa R_0 \sum_{a,b} i \hat{x}^a \tilde{x}^b \Theta(\tilde{x}^b)
  -\frac{1}{2} (q_1-q_0) \sum_{a} \hat{x}^a \hat{x}^a  \nonumber \\ 
&&-\frac{1}{2} q_0 \sum_{a,b} \hat{x}^a \hat{x}^b
-i\sum_{a} \tilde{x}^a \Theta(-\tilde{x}^a)\hat{x}^a -i \nu \sum_{a} \hat{x}^a- E \sum_a \tilde{x}^a \Theta(\tilde{x}^a) -  \hat{p} \sum_a \Theta(\tilde{x}^a) \} \ .\nonumber 
\end{eqnarray}
A particularly efficient way to disentangle the three sums over the replica-replica couplings is to use 
two coupled Gaussian integrals over variables termed $a$ and $b$ echoing the original average over the 
payoff matrices, which yield 
\begin{eqnarray}
\label{appch31Gfac}
G&=&
\ln \int da db \, p_{\tilde{\kappa}}(a,b) 
\prod_a \int \frac{d \tilde{x}^a d \hat{x}^a}{2 \pi} \exp \{ -\frac{1}{2}\sum_{a} (\hat{q}_1+\hat{q}_0) 
\tilde{x}^a \tilde{x}^a  \Theta(\tilde{x}^a)  \nonumber \\
&& +  \kappa (R_1-R_0) \sum_{a} i \hat{x}^a \tilde{x}^a \Theta(\tilde{x}^a) 
  -\frac{1}{2} (q_1-q_0) \sum_{a} \hat{x}^a \hat{x}^a 
+a \sqrt{\hat{q}_0} \sum_a \tilde{x}^a \Theta(\tilde{x}^a) \nonumber \\ 
&&+ ib\sqrt{q_0} \sum_a \hat{x}^a
-i\sum_{a} \tilde{x}^a \Theta(-\tilde{x}^a)\hat{x}^a -i \nu^x \sum_{a} \hat{x}^a- E \sum_a \tilde{x}^a \Theta(\tilde{x}^a) 
-  \hat{p} \sum_a \Theta(\tilde{x}^a) \} \\
&:=& \ln \int da db \, p_{\tilde{\kappa}}(a,b)  \prod_a \int {\cal D}(\tilde{x}^a,\hat{x}^a)  
\ ,\nonumber 
\end{eqnarray}
where $p_{\tilde{\kappa}}(a,b)$ with $\tilde{\kappa}=\frac{\kappa R_0}{\sqrt{q_0 \hat{q}_0}}$ is defined by 
(\ref{ch3pdef}). 
The resulting expression factorizes giving $n$ identical integrals over $\tilde{x}$ and $\hat{x}$, which may 
be easily performed by considering the cases $\tilde{x}<0$ and $\tilde{x}>0$ separately. The limit $n \to 0$ of 
(\ref{partrep2}) may now be taken yielding (\ref{Srep})-(\ref{Srephelp}). 

\section{The stability of the replica-symmetric saddle-point}
\label{appch31rsstab}

In this section we outline the calculation of the eigenvalues of the Hessian matrix of 
(\ref{partrep2}) in order to check if the ansatz (\ref{partrep2}) is locally stable 
against small fluctuations of the order parameters. We focus on the so-called replicon modes  
\cite{AT} and restrict ourselves to the case $\kappa=0$. In this case the Hessian matrix of fluctuations 
of (\ref{partrep2}) around  (\ref{ch3rs}) is given by (\ref{ch3Mdef}). 

The derivatives of $G^x$ and $G^y$ are evaluated at the RS-saddle point. 
Due to symmetry of $G^x$ and $G^y$ under an interchange of the players we have to find the replicon 
eigenvalues of three different submatrices of $M$, beginning with 
$\frac{\partial^2 G}{\partial \hat{q}_{ab} \partial \hat{q}_{cd}}$:  
At the replica-symmetric saddle point there are three different entries in the $\frac{n(n-1)}{2}$ by 
$\frac{n(n-1)}{2}$ matrix of derivatives with respect to the off-diagonal elements of $\hat{q}_{ab}$ with $a>b$. 
These are  
\begin{equation}
\frac{\partial^2 G}{\partial\hat{q}_{ab} \partial\hat{q}_{cd}} = \left\{
\begin{array}{ll}
P_1 & \mbox{for} \;\; a=c, b=d \\
Q_1 & \mbox{for} \;\; \mbox{exactly one pair of indices equal} \\
R_1 & \mbox{for} \;\; a \neq c, b \neq d
\end{array}
 \right. \ ,  
\end{equation}
where
\begin{eqnarray}
P_1&=&\langle x_a^2 x_b^2 \rangle - \langle x_a x_b \rangle \langle x_a x_b\rangle \nonumber \\
Q_1&=&\langle x_a^2 x_b x_c\rangle - \langle x_a x_b \rangle \langle x_a x_c\rangle \nonumber \\
R_1&=&\langle x_a x_b x_c x_d\rangle - \langle x_a x_b \rangle \langle x_c x_d\rangle \ ,
\end{eqnarray}
the pointed brackets denote the normalized averages over 
\begin{eqnarray}
\langle(.)\rangle&=&
\frac{\prod_a \int \frac{d \tilde{x}^a d \hat{x}^a}{2 \pi} \exp \{{\cal L}^x(\{\tilde{x}^a,\hat{x}^a\}) \} (.) }
{\prod_a \int \frac{d \tilde{x}^a d \hat{x}^a}{2 \pi} \exp \{{\cal L}^x(\{\tilde{x}^a,\hat{x}^a\}) \}}
\ ,
\end{eqnarray}
and ${\cal L}^x$ is defined in (\ref{Gdef}) and the order parameters take on their saddle point values. 
In the limit $n \to 0$ the replicon eigenvalue of this matrix equals 
\begin{eqnarray}
\lambda_1&=&P_1-2Q_1+R_1= \int da db \, p_{\tilde{\kappa}}(a,b) \left[\frac{\int {\cal D}(\tilde{x},\hat{x}) \tilde{x}^2 \Theta(\tilde{x})}{\int {\cal D}(\tilde{x},\hat{x})}-\left(\frac{\int {\cal D}(\tilde{x},\hat{x}) \tilde{x} \Theta(\tilde{x})}{\int {\cal D}(\tilde{x},\hat{x})}\right)^2 \right]^2 \nonumber \\
&=& \int da db \, p_{\tilde{\kappa}}(a,b) \left[ \frac{\partial^2}{\partial E^2} \ln L(a,b) \right]^2  
\end{eqnarray}
where ${\cal D}(\tilde{x},\hat{x})$ is defined in (\ref{appch31Gfac}) and $L(a,b)$ is defined in 
(\ref{Gdef}). 

The replicon eigenvalue of $\frac{\partial^2 G}{\partial q_{ab} \partial q_{cd}}$ 
is evaluated in the same fashion. We obtain 
\begin{equation}
\frac{\partial^2 G}{\partial q_{ab} \partial q_{cd}} = \left\{
\begin{array}{ll}
P_2 & \mbox{for} \;\; a=c, b=d \\
Q_2 & \mbox{for} \;\; \mbox{exactly one pair of indices equal} \\
R_2 & \mbox{for} \;\; a \neq c, b \neq d
\end{array}
 \right. \ ,  
\end{equation}
where
\begin{eqnarray}
P_2&=&\langle \hat{x}_a^2 \hat{x}_b^2 \rangle - \langle \hat{x}_a \hat{x}_b \rangle \langle \hat{x}_a \hat{x}_b\rangle \nonumber \\
Q_2&=&\langle \hat{x}_a^2 \hat{x}_b \hat{x}_c\rangle - \langle \hat{x}_a \hat{x}_b \rangle \langle \hat{x}_a \hat{x}_c\rangle \nonumber \\
R_2&=&\langle \hat{x}_a \hat{x}_b \hat{x}_c \hat{x}_d\rangle - \langle \hat{x}_a \hat{x}_b \rangle \langle \hat{x}_c \hat{x}_d\rangle \ ,
\end{eqnarray}
In the limit $n \to 0$ the replicon eigenvalue of this matrix equals 
\begin{eqnarray}
\lambda_2&=&P_2-2Q_2+R_2= \int da db \, p_{\tilde{\kappa}}(a,b) \left[\frac{\int {\cal D}(\tilde{x},\hat{x}) \hat{x}^2 }{\int {\cal D}(\tilde{x},\hat{x})}-\left(\frac{\int {\cal D}(\tilde{x},\hat{x}) \hat{x}} {\int {\cal D}(\tilde{x},\hat{x})}\right)^2 \right]^2 \nonumber \\
&=& \int da db \, p_{\tilde{\kappa}}(a,b) \left[ \frac{\partial^2}{\partial \nu^2} \ln L(a,b) \right]^2 \ . 
\end{eqnarray}

The matrix $\frac{\partial^2 G}{\partial q_{ab} \partial \hat{q}_{cd}}$ also 
consists of three different entries. These are  
\begin{equation}
\frac{\partial^2 G}{\partial\hat{q}_{ab} \partial q_{cd}} = \left\{
\begin{array}{ll}
P_3 & \mbox{for} \;\; a=c, b=d \\
Q_3 & \mbox{for} \;\; \mbox{exactly one pair of indices equal} \\
R_3 & \mbox{for} \;\; a \neq c, b \neq d
\end{array}
 \right. \ ,  
\end{equation}
where
\begin{eqnarray}
P_3&=&-\langle x_a x_b \hat{x}_a \hat{x}_b \rangle + \langle x_a x_b \rangle \langle \hat{x}_a \hat{x}_b\rangle \nonumber \\
Q_3&=&-\langle  x_a x_b \hat{x}_a \hat{x}_c\rangle + \langle x_a x_b \rangle \langle \hat{x}_a \hat{x}_c\rangle \nonumber \\
R_3&=&-\langle x_a x_b \hat{x}_c \hat{x}_d\rangle + \langle x_a x_b \rangle \langle \hat{x}_c \hat{x}_d\rangle \ .
\end{eqnarray}
In the limit $n \to 0$ the replicon eigenvalue of this matrix equals 
\begin{eqnarray}
\lambda_3&=&P_3-2Q_3+R_3= -\int da db \, p_{\tilde{\kappa}}(a,b) \left[\frac{\int {\cal D}(\tilde{x},\hat{x}) \tilde{x}\hat{x} \Theta(\tilde{x})}{\int {\cal D}(\tilde{x},\hat{x})}-\frac{\int {\cal D}(\tilde{x},\hat{x}) \tilde{x} \Theta(\tilde{x})}{\int {\cal D}(\tilde{x},\hat{x})}
\frac{\int {\cal D}(\tilde{x},\hat{x}) \hat{x}}{\int {\cal D}(\tilde{x},\hat{x})} \right]^2 \nonumber \\
&=& \int da db \, p_{\tilde{\kappa}}(a,b) \left[ \frac{\partial^2}{\partial \nu \partial E} \ln L(a,b) \right]^2 \ . 
\end{eqnarray}

Since the replicon eigenvectors of these three matrices are parallel, the eigenvalues of (\ref{ch3Mdef}) 
are those of the matrix 
\begin{equation}
  \left(
  \begin{array}{llll}
\lambda_2 & -1 & 0 & \lambda_3    \\
  -1 &\lambda_1 & \lambda_3 &0\\
  0 &\lambda_3& \lambda_2 & -1 \\
 \lambda_3& 0 &-1& \lambda_1
  \end{array} \right) 
\end{equation}
and we denote the coefficients of replicon-fluctuations as 
$\delta q^x,\delta \hat{q}^x, \delta q^y,\delta \hat{q}^y$.   
In order to determine the criterion for local stability of the RS-saddle point we 
first eliminate the fluctuations in the conjugate order 
parameters $\delta\hat{q}^x$ and $\delta\hat{q}^y$ near the saddle point. From 
$\partial S / \partial \delta \hat{q}^y=0$ and 
$\partial S / \partial \delta \hat{q}^y=0$ one obtains 
$\delta \hat{q}^x=\frac{1}{\lambda_2}(\delta q^x- \lambda_3 \delta q^y)$ and 
$\delta \hat{q}^y=\frac{1}{\lambda_2}(\delta q^y- \lambda_3 \delta q^x)$ respectively. 

This allows us to write the matrix of replicon fluctuations in terms of $\delta q^x$ and $\delta q^y$ 
only yielding after some algebra 
\begin{equation}
 S=S_{RS} + \frac{1}{2} (\delta q^x \delta q^y) M' 
\left( \begin{array}{ll} \delta q^x \\ \delta q^y \end{array} \right)  + {\cal O}(\delta^3) 
\end{equation} 
with 
\begin{equation}
M'=
\frac{1}{\lambda_2}
  \left(
  \begin{array}{ll}
\lambda_1 \lambda_2 -\lambda_3^2-1 & 2 \lambda_3 \\
2 \lambda_3 & \lambda_1 \lambda_2 -\lambda_3^2-1 
  \end{array} \right) \ .
\end{equation}
Since the integrals over the variables $q^x$ and $q^y$ are now over a real function, the criterion that 
the RS ansatz (\ref{partrep2}) is locally stable is that both eigenvalues of $M'$ are 
negative giving 
\begin{eqnarray}
\frac{1}{\lambda_2}(\lambda_1 \lambda_2 -(\lambda_3-1)^2) &<& 0 \nonumber \\
\frac{1}{\lambda_2}(\lambda_1 \lambda_2 -(\lambda_3+1)^2) &<& 0	\ .
\end{eqnarray}


\begin{thebibliography}{10}
\bibitem{probskappa}
At non-zero $\kappa$ cross-terms e.g of the form $\partial^2 \uppercase{S}/
  \partial q^y_{ab} \partial \uppercase{R}^y_{cd}= \langle x_a x_b i \hat{x}_c x_d
  \rangle$ arise. \uppercase{A}t the \uppercase{RS} saddle-point these matrices
  consist of four different elements and the standard analysis of \cite{AT} is
  no longer applicable.
\end{thebibliography}
\end{document}